\documentstyle[epsfig,12pt]{article}
\newcommand{\case}[2]{\mbox{\footnotesize $\displaystyle \frac{#1}{#2}$}}
\begin{document}
{\Large\bf Leptonic and semileptonic decays of heavy mesons}\\[2mm]
{\it C. D. Roberts$\,^a$, M. A. Ivanov$\,^b$, Yu. L. Kalinovsky$\,^c$ and P. Maris$\,^a$}\\[2mm]
{\small $^a$Physics Division 203, Argonne National Laboratory, Argonne IL
60439-4843, USA\\
$^b$Bogoliubov Laboratory of Theoretical Physics, 
JINR, 141980 Dubna, Russia\\
$^c$Laboratory of Computing Techniques and Automation, 
JINR, 141980 Dubna, Russia}\\[2mm]
The leptonic decay of a pseudoscalar meson with total momentum $P$ is
described by
\begin{eqnarray}
\lefteqn{\langle 0 | \bar q_{f_2} \gamma_\mu \gamma_5 q_{f_1}| \Phi_M(P)\rangle 
:= }\nonumber \\
&& \label{lepdecay}
f_M P_\mu = N_c \int\frac{d^4 k}{(2\pi)^4} 
{\rm tr}_D\left[\gamma_5 \gamma_\mu S _{f_1}(k) 
\Gamma _M(k;P)  S _{f_2}(k - P)\right]\,,
\end{eqnarray}
which defines the leptonic decay constant, $f_M$.  (With this normalisation,
$f_\pi\simeq 131\,$MeV.) In (\ref{lepdecay}), $S_f$ is the dressed quark
propagator and $\Gamma _M(k;P)$ is the Bethe-Salpeter amplitude for the bound
state; $M$ labels the meson whose flavour content is made explicit by the
quark flavour labels, $f_i$.  For example, for the $B^-$-meson: $f_1 = u$ and
$f_2=b$.

The calculation of $f_M$ requires a knowledge of $S_f$ and $\Gamma_M$.  The
dressed-quark propagator has the general form $S_f(p)=1/[i\gamma\cdot p
A_f(p^2) + B_f(p^2)]$: a bare-quark is described by $A(p^2)\equiv 1$ and
$B(p^2)=m$, where $m$ is the current-quark mass.  As described in
Ref.~[\ref{refa}], it is a characteristic of QCD elucidated in
Dyson-Schwinger equation studies that for light-quarks; i.e., $u$-, $d$- and
$s$-quarks, $A_f(p^2)$ and particularly $B_f(p^2)$ have a strong
momentum-dependence for $p^2<1\,$GeV$^2$.  This momentum-dependence is
nonperturbative in origin.

For the $b$-quark, however, the momentum-dependence of $A_b(p^2)$ and
$B_b(p^2)$ for all spacelike-$p^2$ is weak and mainly perturbative in origin.
This suggests that, in phenomenological applications, it is a good
approximation to write the dressed-$b$-quark propagator as
\begin{equation}
S_b(p) = \frac{1}{i\gamma\cdot p + \hat M_b}\,,
\label{sb}
\end{equation}
where $\hat M_b$ is approximately the Euclidean constituent-quark
mass$\,^{\ref{refa}}$.  As observed in Ref.~[\ref{refb}], this is the origin
of ``heavy-quark symmetry'' in the Dyson-Schwinger equation [DSE] approach.
For the $c$-quark, $A_c(p^2)$ and $B_c(p^2)$ have a stronger
momentum-dependence.  Hence representing $S_c$ analogously to (\ref{sb}) is
only, at best, a first, exploratory step in the study of heavy meson
properties.  To proceed we write the heavy-meson total-momentum as $P:= (\hat
M_{f_Q} + E)\,v_\mu$, where $E= M_H - \hat M_{f_Q}$ and $v^2=-1$.  It follows
that the heavy-quark propagator in (\ref{lepdecay}) becomes
\begin{equation}
\label{heavy}
S_{f_2 = c,b}(k-P)= -\,\frac{1}{2}\,\frac{1 + i \gamma\cdot v}{k\cdot v + E}
        + {\rm O}\left(\frac{|k|}{\hat M_{f_Q}},
                \frac{E}{\hat M_{f_Q}}\right)\,.
\end{equation}

The Bethe-Salpeter amplitude, $\Gamma_M(k;P)$, in (\ref{lepdecay}) is a
function of the light-quark's momentum, $k$.  It can be obtained as the
solution of a Bethe-Salpeter equation$\,^{\ref{refa}}$.  These studies have
not yet been completed hence herein we employ the Ansatz
\begin{equation}
\label{Gamma}
\Gamma_{B,D}(k;P)  = 
        \gamma_5 \left( 1 - \case{1}{2} i\gamma\cdot v \right)
                \frac{ \varphi(k^2)}{ {\cal N}_{B,D}}\,,
\end{equation}
whose Dirac structure is motivated by Ref.~[\ref{refc}].  Here ${\cal N}_H$
is the canonical normalisation constant for the Bethe-Salpeter amplitude.  In
this study we interpret an insensitivity of our results to details of the
form of $\varphi(k^2)$ as indicating that they are robust.

Using (\ref{heavy})
\begin{equation}
\label{norm}
 \frac{1}{M_H \kappa_f^2}  :={\cal N}_H^2
= \frac{1}{M_H}\,\frac{N_c}{32 \pi^2}
\int_0^\infty du \,
\varphi(z)^2\,
\left({\sigma}_S^f(z) + \sqrt{u} \,{\sigma}_V^f(z)\right)\,,
z= u - 2 E \surd u\,,
\end{equation}
where we have introduced the notation 
$
S_f(k) = -i\gamma\cdot k \sigma_V^f(k^2) + \sigma_S^f(k^2)
$
for the light-quark propagator.  Substituting (\ref{heavy}) and (\ref{Gamma})
into (\ref{lepdecay}), and using (\ref{norm}), we obtain
\begin{equation}
f_{H} = \frac{\kappa_f}{ \sqrt{M_{H}}}\,
        \frac{N_c}{8\pi^2}\,
        \int_0^\infty\,du\,
        (\sqrt{u}-E)\,{\varphi}(z)\,
        \left({\sigma}_S^f(z) 
        + \case{1}{2} \sqrt{u} {\sigma}_V^f(z)\right)\,,
\end{equation}
which entails that in the heavy-quark limit, as we have defined it herein, 
\begin{equation}
f_H\,\propto \frac{1}{\sqrt{M_H}}\,.
\end{equation}
Contemporary estimates of $f_D$ and $f_B$ suggest that this scaling relation
is not applicable to experimentally accessible heavy mesons.

In impulse approximation the hadronic matrix element for the $B^0\to D^- \ell
\nu$ decay is$\,^{\ref{refb}}$
\begin{eqnarray}
\lefteqn{\langle D^-(K) | \bar b \gamma_\mu c | B^0(P) \rangle
:= f_+(t) (K+P)_\mu - f_-(t) (K-P)_\mu} \nonumber \\
&& \label{semlep}
= N_c \int\frac{d^4 \ell}{(2\pi)^4}
{\rm tr}_D\left[\bar{\Gamma}_{D^-}\left(\ell; - K\right)
 S_d(\ell ) \Gamma_{B^0}\left(\ell;P\right) 
 S_b(\ell - \eta P)\,i\gamma_\mu\, S_c(\ell- \eta K )\right]\,,
\end{eqnarray}
where $t=-(P-K)^2$ and $\bar \Gamma_{B^0,D^-}(k;P)^{\rm T} =
C^\dagger\Gamma_{B^0,D^-}(-k;P)C$, with $C=\gamma_2\gamma_4$.  In
(\ref{semlep}) we have used the fact that in the heavy-quark limit the vector
piece of the dressed-quark-W-boson vertex is $V_\mu^{bc}=\gamma_\mu$.
Substituting (\ref{heavy}) and (\ref{Gamma}) into (\ref{semlep}) we obtain
\begin{eqnarray}
f_\pm (t) & = &\case{1}{2} \,\frac{M_D \pm M_B}{\sqrt{M_D M_B}} 
        \,{\xi}(w)\,,\\
\label{xiw}
{\xi}(w) & = &\kappa_d^2 \frac{N_c}{32\pi^2}\int_0^1 d\tau\,\frac{1}{W}\,
\int_0^\infty du\, {\varphi}(z_W)^2\,
        \left({\sigma}_S^d(z_W) + \sqrt{\frac{u}{W}}\,
{\sigma}_V^d(z_W)\right)\,,
\end{eqnarray}
with $W= 1 + 2 \tau (1-\tau) (w-1)$ and $z_W= u - 2 E \sqrt{u/W}$.  In
(\ref{xiw}), $w = \frac{M_B^2 + M_D^2 - t}{2 M_B M_D} = v_B \cdot v_D$ and
the physically accessible region is $1.0<w<1.6$.  The canonical normalisation
of the Bethe-Salpeter amplitude, (\ref{norm}), ensures that $\xi(w=1)=1$.

One quantity characterising the function $\xi(w)$ is its slope at $w=1$, the
point of minimal heavy meson recoil: $\rho^2:=
-\left.\xi^\prime(w)\right|_{w=1}$.  It follows from (\ref{xiw}) that
$\rho^2\geq 1/3$ for any $\varphi(z)$ and $\sigma_{V/S}^f(z)$ non-negative,
non-increasing, convex-up functions of their argument, which includes
$\varphi=\,$constant and a free-particle propagator.

At this point the calculation of the leptonic decay constants and $\xi(w)$
wants only the specification of the light-quark propagators and the function
$\varphi(k^2)$.  The light-quark propagators have been fixed in
Ref.~[\ref{refd}]:
\begin{eqnarray}
\label{SSM}
\bar\sigma^f_S(x)  & =  & 
        2 \bar m_f {\cal F}(2 (x + \bar m_f^2))
        + {\cal F}(b_1 x) {\cal F}(b_3 x) 
        \left( b^f_0 + b^f_2 {\cal F}(\epsilon x)\right)\\
\label{SVM}
\bar\sigma^f_V(x) & = & \frac{2 (x+\bar m_f^2) -1 
                + e^{-2 (x+\bar m_f^2)}}{2 (x+\bar m_f^2)^2},
\end{eqnarray}
where ${\cal F}(y)\equiv (1-{\rm e}^{-y})/y$, $x=p^2/(2 D)$ and:
$\bar\sigma_V^f(x) = 2 D\,\sigma_V^f(p^2)$; $\bar\sigma_S^f(x) = \sqrt{2
D}\,\sigma_S^f(p^2)$; $\bar m_f$ = $m_f/\sqrt{2 D}$, with $D$ a mass scale.
This form is motivated by extensive studies$\,^{\ref{DSErev}}$ of the DSE for
the dressed-quark propagator and combines the effects of confinement and
dynamical chiral symmetry breaking with free-particle behaviour at large
spacelike-$p^2$.  The parameters $\bar m_f$, $b_{0\ldots 3}^f$ in
(\protect\ref{SSM}), (\protect\ref{SVM}) were determined in a $\chi^2$-fit to
a range of light-hadron observables, which is described in Ref.~[\ref{refd}]
and leads to the values in (\ref{tableA})
\begin{equation}
\label{tableA} 
\begin{array}{cccccc}
 u:  & 0.00897 & 0.131 & 2.90 & 0.603 & 0.185 \\
 s:  & 0.224   & 0.105 & \underline{2.90} & 0.740 & \underline{0.185}
\end{array}
\end{equation}
The values of $b_{1,3}^s$ are underlined to indicate that the constraints
$b_{1,3}^s=b_{1,3}^u$ were imposed in the fitting.  The scale parameter
$D=0.160\,$GeV$^2$.

We consider the following four forms for $\varphi(k^2)$:
\begin{equation}
\label{formsphi}
\begin{array}{cccccccccc}
\varphi_A(k^2) && &\varphi_B(k^2) & & &\varphi_C(k^2) & &&
\varphi_D(k^2) \\\hline
\displaystyle \exp\left(-\frac{k^2}{\Lambda^2}\right)
& &&\displaystyle  \frac{\Lambda^2}{k^2+\Lambda^2}
& &&\displaystyle  \rule{0mm}{10mm}
\left(\frac{\Lambda^2}{k^2+\Lambda^2} \right)^2
& &&\displaystyle  \theta\left(1 - \frac{k^2}{\Lambda^2}\right)
\end{array}
\end{equation}
and with this we have 2 parameters in our study: the ``binding energy'', $E:=
M_H - \hat M_{f_Q}$, and the width, $\Lambda$, of the heavy meson
Bethe-Salpeter amplitude.
\begin{figure}[t]
\vspace*{-5mm}

\epsfig{figure=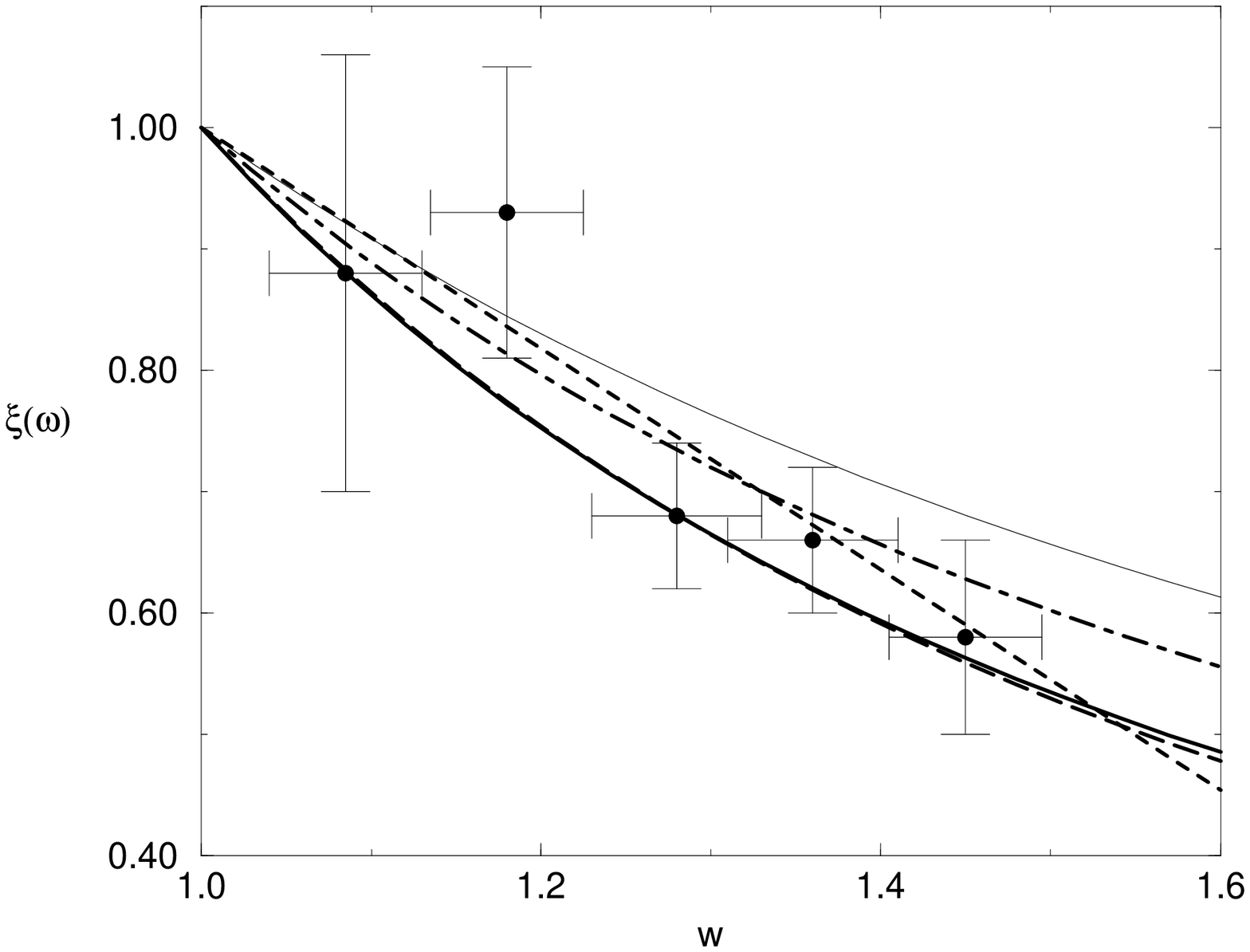,height=7.9cm} \vspace*{-72mm}

\hspace*{90mm}
\parbox{70mm}{\small Fig.~1.~Experiment: points, Ref.~[\protect\ref{argus}];
dashed line, (\protect\ref{CESRfit}); short-dashed line, the linear
fit$\,^{\protect\ref{CESR96}}$ $\xi(w) = 1 - \rho^2 (w - 1)$, $\rho^2 = 0.91
\pm 0.15 \pm 0.06$.  Our calculations using $\varphi_{A-C}$ from
(\protect\ref{results}) are represented by the solid line with the dot-dash
line being the result we obtain assuming a point-like heavy-meson,
$\varphi_D$.  Importantly, there is significant curvature in each case, which
is a manifestation of the role played by the light-quarks. The light solid
line is described in Fig.~2.}
\vspace*{5mm}
\end{figure}

Now we ask the question: ``Is the heavy-quark limit of the DSE framework
capable of describing heavy meson observables?''.  To answer this we perform
a $\chi^2$-fit of $(E,\Lambda)$ to the following parametrisation of the
experimental data$\,^{\ref{CESR96}}$ on $\xi(w)$:
\begin{equation}
\label{CESRfit}
\xi(w)=\frac{2}{w+1} \exp\left((1 - 2 \rho^2)\frac{w-1}{w+1}\right)\,,\;
        \rho^2 = 1.53\pm 0.36 \pm 0.14\,,
\end{equation}
to $f_D = 0.216 \pm 0.015\,$GeV and $f_B = 0.206 \pm 0.030\,$GeV, which, in
the absence of experimental data, is our weighted average of lattice-QCD
results.  Using $M_D= 1.87\,$GeV, $M_{D_s}= 1.97\,$GeV and $M_B=5.27\,$GeV,
we obtain the results presented in Fig.~1 and (\ref{results}), energies in
GeV and $\rho^2$ dimensionless.

\begin{equation}
\label{results}
\begin{array}{lc|cll|cllll}
  &&&  E   &  \Lambda  &&  f_D  &  f_{D_s} &  f_B     &  \rho^2\\\hline
A &&& 0.640 & 1.03  && 0.227  &  0.245  &  0.135  &  1.55 \\
B &&& 0.567 & 0.843 && 0.227  &  0.239  &  0.135  &  1.56 \\
C &&& 0.612 & 1.32  && 0.227  &  0.242  &  0.135  &  1.55 \\
D &&& 0.643 & 1.02  && 0.272  &  0.296  &  0.162  &  1.21 
\end{array}
\end{equation}
Clearly, a good description is possible.  The fitted values of $E$ are
consistent with contemporary estimates of this binding energy in
Bethe-Salpeter equation studies$\,^{\ref{mj92}}$ and the values of $\Lambda$
indicates that the heavy meson occupies a spacetime volume of only
$4\,$-$\,20$\% that of the pion.  We observe that $f_{D_s}/f_D = f_{B_s}/f_B
\approx 1.07$.  Comparing this with the value expected in the heavy-quark
limit: $\sqrt{M_D/M_{D_s}}= 0.97$ and $\sqrt{M_B/M_{B_s}}= 0.99$, illustrates
the influence that light-quarks have on real heavy-meson observables.

\begin{figure}[h]
\vspace*{-6mm}

\epsfig{figure=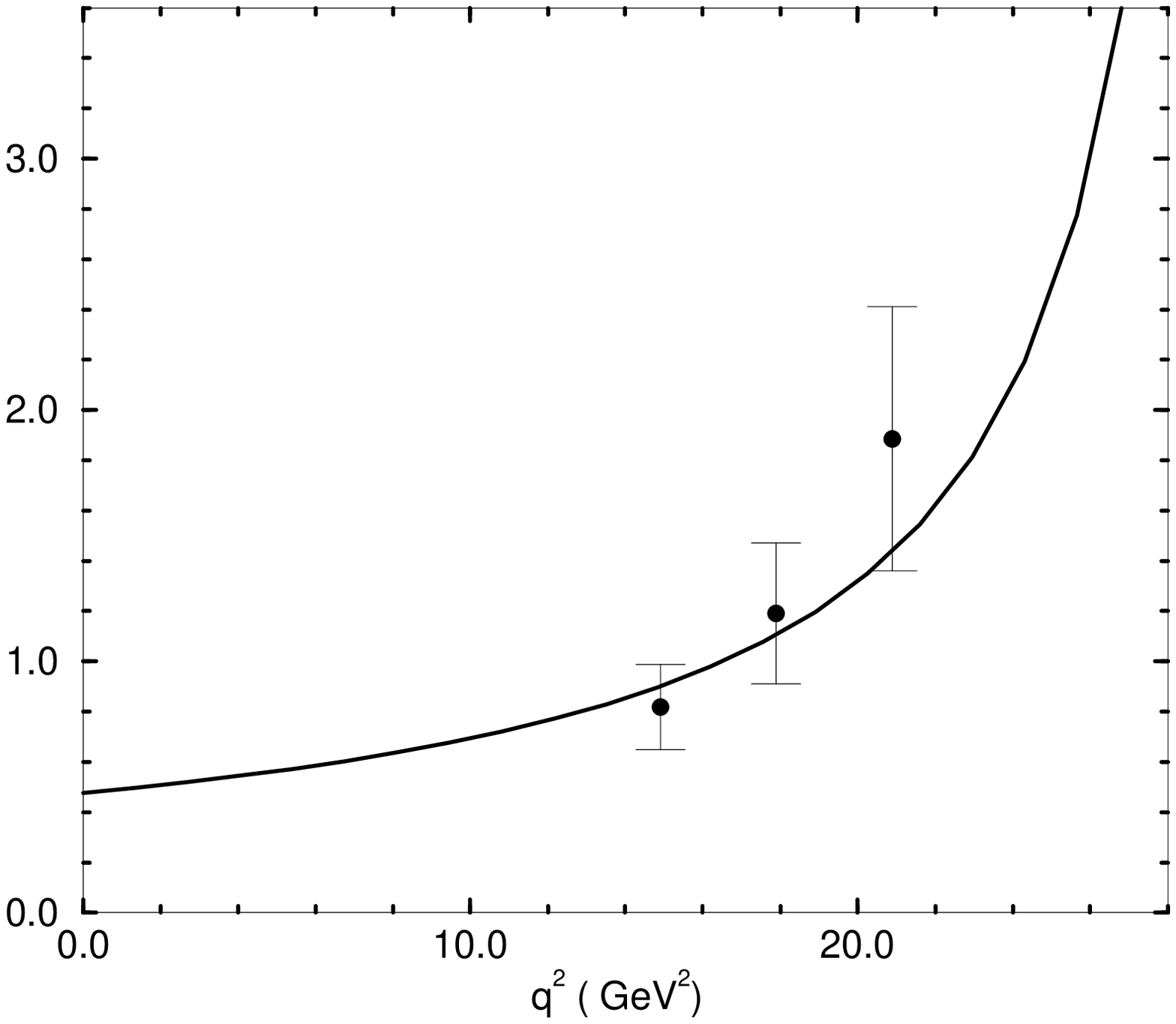,height=7.9cm} \vspace*{-73mm}

\hspace*{90mm}
\parbox{70mm}{\small Fig.~2.~Calculated form of $f_+(q^2)$ for the decay
$B\to \pi e \nu$ using $\varphi_A$, with $E=0.47$ and $\Lambda= 1.1\,$GeV,
and $\Gamma_\pi$ from Ref.~[\protect\ref{refd}]: $f_+(0)=0.48$.  This gives a
branching ratio of $2.3 \times 10^{-4}$ to be compared with the experimental
value$\,^{\protect\ref{cleo96}}$ of $(1.8\pm 0.4\pm 0.3 \pm 0.2)\times
10^{-4}$.  With this $\varphi_A$: $f_D= 0.224$, $f_{D_s}=0.241$,
$f_B=0.133\,$ and $f_{B_s}=0.146\,$GeV.  $\xi(w)$ is plotted as the thin solid
line in Fig.~1, for which $\rho^2=1.0$.  Requiring a simultantaneous fit
reduces $\rho^2$ and increases $\xi(w=1.6)$.  The data points are the results
of the lattice simulations in Ref.~[\protect\ref{refe}].}
\vspace*{0.6\baselineskip}
\end{figure}
We are currently applying the formalism described herein to the simultaneous
calculation of leptonic and heavy-to-heavy and heavy-to-light semileptonic
decays.  Our framework allows the calculation of each form factor at all
$q^2$.  The light-quark degrees of freedom are particularly important in
heavy-to-light semileptonic decays, which probe the structure of the
final-state light-meson Bethe-Salpeter amplitude and are inaccessible in
heavy-quark effective theory.  A uniformly good description of all these
decays requires a refitting of the two parameters $E$ and $\Lambda$.  We
illustrate what is possible in Fig.~2.  

{\bf Acknowledgments}.  The work of P.M. and C.D.R. was supported by the US
Department of Energy, Nuclear Physics Division, under contract number
W-31-109-ENG-38, and benefited from the resources of the National Energy
Research Supercomputer Center.

{\bf References}.\\[-1.6\baselineskip]
\begin{enumerate}
\item \label{refa} P. Maris and C. D. Roberts. ``Differences between heavy
and light quarks'', these\\[-0.2\baselineskip]
proceedings, e-print nucl-th/9710062.\vspace*{-0.67\baselineskip} 
\item \label{refb} M. A. Ivanov {\it et al}., e-print nucl-th/9704039, to
appear in Phys. Lett. B.\vspace*{-0.67\baselineskip}
\item \label{refc} C. J. Burden {\it et al}., Phys. Rev. C {\bf 55} (1997)
2649. \vspace*{-0.67\baselineskip}  
\item \label{refd} C. J. Burden, C. D. Roberts and M. J. Thomson, Phys. Lett. B
{\bf 371} (1996) 163.\vspace*{-0.67\baselineskip}  
\item \label{DSErev} C. D. Roberts and A. G. Williams,
Prog. Part. Nucl. Phys. {\bf 33} (1994) 477.\vspace*{-0.67\baselineskip}  
\item \label{CESR96} CLEO Coll. (J.E. Duboscq {\it et al}.),
Phys. Rev. Lett. {\bf 76} (1996) 3899.\vspace*{-0.67\baselineskip}
\item \label{argus} ARGUS Collaboration, Z. Phys. C {\bf 57} (1993)
249.\vspace*{-0.67\baselineskip} 
\item \label{mj92} H. J. Munczek and P. Jain, Phys. Rev. D {\bf 46} (1992)
438.\vspace*{-0.67\baselineskip}
\item \label{cleo96} CLEO Coll. (J. P. Alexander {\it et al}.),
Phys. Rev. Lett. {\bf 77} (1996) 5000.\vspace*{-0.67\baselineskip}
\item \label{refe} UKQCD Coll. (D. R. Burford {\it et al}.), Nucl. Phys. B
{\bf 447} (1995) 425.\vspace*{-0.5\baselineskip}
\end{enumerate}

\end{document}